\documentclass{article}
  \usepackage{amsfonts,amsmath,amssymb,hhline,epsfig,dcolumn,bbm}
 \usepackage{natbib}
\usepackage[nolists]{endfloat}
\usepackage{vmargin,tabularx,setspace, lineno}
\usepackage[latin1]{inputenc}
\usepackage[T1]{fontenc}

\newcolumntype{.}{D{.}{.}{-1}}

\newcommand{\y}{\mathbf{y}} 

\date{}

\title{Extension of the SAEM algorithm for nonlinear mixed models with two levels of random effects}

\author{Xavi\`ere Panhard$^{1*}$ Adeline Samson$^{2}$}
\begin{document}

\maketitle
$^1$ INSERM U738, Paris, France; University Paris 7, UFR de M\'edecine, Paris, France 

$^2$ Laboratoire MAP5, UMR CNRS 8145, University Paris 5, France

* Corresponding author: INSERM U738, Universit\'e Paris 7, UFR de M\'edecine, 16 rue  Huchard, 75018 Paris, France

\begin{abstract}

This article focuses on parameter estimation of  multi-levels nonlinear mixed effects models (MNLMEMs).
These models are used to analyze data presenting multiple hierarchical levels of grouping (cluster data, clinical trials with several observation periods,...).
 The variability of the individual parameters of the regression function is thus decomposed as a between-subject  variability and higher levels of variability (for example within-subject variability). We propose maximum likelihood estimates of  parameters of those MNLMEMs with two levels of random effects, using an extension of the SAEM-MCMC algorithm. The extended SAEM algorithm is  split into an explicit direct EM algorithm and a stochastic EM part. Compared to the original algorithm,  additional sufficient statistics have to be approximated by relying on  the conditional distribution of the second level of random effects. 
This estimation method is  evaluated on pharmacokinetic cross-over simulated trials, mimicking theophyllin concentration data.  Results   obtained on those datasets with  either the  SAEM algorithm or  the FOCE algorithm (implemented in the nlme function of R software) are compared: biases and RMSEs of almost all the SAEM estimates are smaller  than  the FOCE ones. 
Finally, we apply the extended SAEM algorithm to analyze the pharmacokinetic interaction of
tenofovir on atazanavir, a novel protease inhibitor, from the ANRS 107-Puzzle 2 study. 
A significant decrease of the area under the curve of  atazanavir is found in patients receiving both treatments.
\end{abstract}

\section{Introduction}
 
The use of non-linear mixed effects models (NLMEMs)  increases in several fields such as agronomy, forestry, clinical trials, population pharmacokinetics (PK) and pharmacodynamics or viral dynamics to model longitudinal data. In some settings, data can present multiple hierarchical levels of grouping, leading to multiple nested levels of variability. For instance, we may study patients that are grouped in medical services that are themselves grouped into hospitals. In this article, we call multilevel non-linear mixed effects models (MNLMEMs) the models that describe such data. MNLMEMs represent a natural extension of models with single variability level, and they have recently been subject to a great deal of attention in statistical literature. In the field of forestry, Hall and Clutter \cite{hal04} analyze longitudinal measures of yield and growth that are measured on each tree within a plot. In the field of agronomy, Rekaya et al. \cite{rek01} consider milk yield data where each cow is observed longitudinally during its first three lactations. Jaffrézic et al. \cite{jaf06} perform genetic analyses of growth measurements in beef cattle acknowledging the fact that several cows come from the same sire. Another example is population PK, where concentration measurements may be taken with several patients over several distinct time intervals, that are often named periods or occasions. That grouping pattern is used for instance in PK cross-over trials.

In NLMEMs with only one level of variability, often corresponding to between-subject variability, the analysis results in the estimation of the fixed effects parameters and of the between-subject variability of the parameters, also called inter-subject variability. When there is more than one level of grouping, the higher levels of variability can be estimated. In the specific case where the second level of grouping corresponds to multiple periods of measurement, this variability is called within-subject variability (or intra-subject variability, or inter-occasion variability), and corresponds to the variation of the individual parameters across the different study periods or units. 
In the context of pharmacokinetics, Karlsson et al. \cite{kar93} demonstrate the importance of modeling this second level of variability in two-levels NLMEMs. They show that neglecting it resulted in biased estimates for the fixed effects. 

The parameter estimation of NLMEMs is not trivial because the likelihood of NLMEMs cannot be expressed in a closed form due to the non-linearity of the regression function in the random effects. 
Therefore, several estimation methods have been proposed.
The First Order Conditional Estimates (FOCE) algorithm performs a first order linearization of the regression function  with respect to the random effects \cite{bea82, lin90}.  The implementation of the FOCE  algorithm in the NONMEM software and in the nlme function of Splus and R enables the estimation of both between- and within-subject variabilities. From our practice, the main drawback of this method is however that it does not always converge when one estimates simultaneously the between- and the within-subject variabilities on several parameters. Furthermore, this linearization-based method cannot be considered as fully established in theory. For instance, Vonesh \cite{von96}  and Ge et al. \cite{ge04} give  examples of  specific designs resulting in inconsistent estimates, such as when the number of observations per subject does not increase faster than the number of subjects or when the variability of random effects is too large. 

Several estimation methods have been proposed as  alternatives to linearization algorithms. 
A common method to handle numerical integration is the adaptative Gaussian quadrature (AGQ) method. 
An estimation algorithm of NLMEM parameters based on this classical AGQ method has been proposed by Pinheiro and Bates \cite{pin95} and is implemented in the SAS procedure NLMIXED \cite{wol93}. 
However, the AGQ method requires a sufficiently large number of quadrature points implying an often slow convergence with very high computational time. Furthermore, two-levels NLMEM can be implemented in the NLMIXED procedure, but to our knowledge, the convergence is difficult to obtain in practice \cite{jaf06}. Improvements upon this method are thus needed. 
The second alternative to linearization is the use of the Expectation-Maximization (EM) algorithm \cite{dem77} in order to estimate models with missing or non-observed data such as random effects.   
Because of the nonlinearity of the model, stochastic versions of the EM algorithm have been proposed.  
Wei et al. \cite{wei90}; Walker \cite{wal96} and Wu \cite{wu04} propose MCEM algorithms, with  a Monte-Carlo approximation of   the E-step. 
However the MCEM algorithm may have computational problems (i.e slow or even non convergence). 
 As an alternative to address  computational problems, a stochastic approximation version of EM (SAEM) has been proposed in \cite{del99,kuh05}, which requires the simulation of only one realization of the missing data for each iteration, substantially reducing the computation time. 
Kuhn and Lavielle \cite{kuh05} propose to combine the SAEM algorithm with a Monte-Carlo Markov Chain (MCMC) procedure adapted to the NLMEMs, and  prove that the resulting estimates are convergent and consistent. 

To date, none of the EM-based algorithms are directly applicable to the case of multilevel NLMEMs and have to be adapted. The objective of this paper is to extend the SAEM algorithm to MNLMEMs with two levels of variability: both E and M steps need to be adapted to integrate higher levels of random effects. We also propose estimates of the likelihood and of the Fisher information matrix.
We  evaluate this algorithm on a PK example, more precisely a two-periods one-sequence cross-over trials simulated mimicking theophyllin concentration data \cite{pin95}. 
We also apply the SAEM algorithm to the PK interaction of two HIV molecules (tenofovir and atazanavir) from a PK substudy of the ANRS 107-Puzzle 2 trial.
After describing the model and notations (section  \ref{model}),  section \ref{saem} describes the SAEM algorithm. 
Section \ref{simu} reports the simulation study and its results. In Section \ref{ataza}, we study the
PK interaction of tenofovir on atazanavir in HIV patients. 
Section \ref{disc} concludes the article with some discussion.

\section{Models}
\label{model}
Let us denote $y_{ijk}$  the observation in unit $k$ $(k=1,\ldots,K)$ for subject $i$ $(i=1,\cdots,n)$ and at time $t_{ijk}$ $(j=1,\cdots,n_{ik})$. For instance, the different units can be the  different periods in the case of PK trials, or the different parents in the case of genetic analyses. 
We assume, as a known fact, two nonlinear functions $f$ and $g$ such that the two-levels non-linear mixed effects model linking observations to sampling times can be written as:
\begin{eqnarray*}
y_{ijk}&=&f(t_{ijk},\phi_{ik})+g(t_{ijk},\phi_{ik})\varepsilon_{ijk},\\
\varepsilon_{ijk}&\sim&\mathcal{N}(0,\sigma^2),
\end{eqnarray*}
where $\phi_{ik}$ is the $p$-vector of the parameters of subject $i$ for unit $k$ and
$\varepsilon_{ijk}$ is the measurement error. We hypothesize that the errors
$\varepsilon_{ijk}$ given $\phi_{ik}$ are mutually independent. 
We assume that the individual parameters $\phi_{ik}$ are  random vectors and that
for each unit $k$, $\phi_{ik}$ can be broken into:
\begin{eqnarray}
\phi_{ik}&=&\mu+\beta_{k}+b_{i}+c_{ik}, \label{eq:ind_par}\\
b_i&\sim&\mathcal{N}(0,\Omega),\nonumber\\
c_{ik}&\sim&\mathcal{N}(0,\Psi),\nonumber
\end{eqnarray}
where $\mu+\beta_{k}$ is the mean value for unit $k$,  $b_{i}$ is the random effect of size $p$ of subject $i$, and  $c_{ik}$  is the random effect of size $p$  of subject $i$ and unit $k$. To ensure the identifiability of the parameters, we assume that $\beta_1=0$, ie $\mu$ is the mean of the first unit and $\beta_k$ represents the difference (or effect) of the $k$th unit in comparison to this first unit. The random effects $(b_{i})$ and $(c_{ik})$  are assumed to be mutually independent. The total variance of the parameters is thus broken into a between-subject variance $\Omega$ and a within-subject variance $\Psi$. 
   Finally, the  individual parameters $pK$-vector $\phi_i=(\phi_{i1},\ldots,\phi_{iK})$ of subject $i$ is distributed with a Gaussian distribution with mean vector $(\mu,\mu +\beta_2, \ldots,\mu+\beta_{K})$ and a $pK\times pK$ covariance matrix $\Gamma$ equal to 
\begin{equation*}
\Gamma=\left(
\begin{array}{ccccc}
\Omega +\Psi & \Omega &\ldots &\Omega\\
\Omega & \Omega  +\Psi&\ddots& \vdots\\
\vdots & \ddots&\ddots&\Omega\\
\Omega&\ldots&\Omega&\Omega  +\Psi
\end{array}
\right).
\end{equation*}

Let $\theta=(\mu,\beta,\Omega,\Psi,\sigma^2)$, the vector of all the parameters of the model where $\beta$ denotes the vector of unit effect $\beta=(\beta_1,\ldots,\beta_K)$. The aim of this paper is to propose an estimation of $\theta$ by maximizing the likelihood of the observations $\y=(y_{ijk})_{ijk}$. 

Let us denote $\tilde{b}_i:=\mu+b_{i}$. 
The likelihood of $\y$ can be written as:
\begin{eqnarray*}
 p(\y;\theta)&=&\int p(\y,\phi,\tilde{b};\theta) d(\phi,\tilde{b})
 \end{eqnarray*}
where $p(\y,\phi,\tilde{b};\theta)$ is the likelihood of the complete data $(\y,\phi,\tilde{b})$, with $\phi=(\phi_{ik})_{i=1,\ldots,n, k=1,\ldots,K}$ and $\tilde{b}=(\tilde{b}_1,\ldots,\tilde{b}_n)$. 
Because of the nonlinearity of the regression function $f$ with respect to the random effects $\phi_{ik}$, the likelihood has no closed form. 
Therefore, the  maximization of the likelihood in $\theta$, $\theta \in \Theta$, is a complex problem. We propose to use a stochastic version of the EM algorithm, which is presented in detail in the next section.

\section{Estimation algorithm}\label{saem}
\subsection{The SAEM algorithm}
The EM algorithm introduced by Dempster et al. \cite{dem77} is a classical approach to estimate parameters of models with non-observed or incomplete data. 
In two-levels NLMEMs, the non-observed data are the individual parameters $(\phi,\tilde{b})$ and the complete data of the model is  $(\y,\phi,\tilde{b})$.
Let us denote $L_c(\y,\phi,\tilde{b};\theta)=\log p(\y,\phi,\tilde{b};\theta) $  the log-likelihood of the complete data. 
The principle of the iterative EM algorithm is to maximize the function $Q(\theta|\theta')=E(L_c(\y,\phi,\tilde{b};\theta)|\y;\theta')$ where the expectation is the conditional expectation under the posterior distribution $p(\phi,\tilde{b}|y;\theta')$, the maximization of $Q$ being often easier than the direct maximization of the observed data log-likelihood. 
Each iteration of the EM algorithm is computed through  two steps: the Expectation step (E-step) and the Maximization step (M-step). 
At the $\ell^{th}$ iteration of the algorithm, the E  step is the
evaluation of  $Q(\theta \, \vert \,\widehat{\theta}_\ell)$, while the M step
updates $\widehat{\theta}_\ell$ by maximizing $Q(\theta \, \vert \,\widehat{\theta}_\ell)$.

Let us show that the function $Q$ can be reduced in the case of a MNLMEM. 
First, let us note that  as $p(\tilde{b}|\y,\phi;\theta)=p(\tilde{b}|\phi;\theta)$, by application of the Bayes theorem we have:
\begin{equation}\label{bayes}
p(\phi,\tilde{b}|\y;\theta)=p(\tilde{b}|\phi;\theta)p(\phi|\y;\theta)=\prod_{i=1}^n p(\tilde{b}_i|\phi_i;\theta)p(\phi_i|\y_i;\theta).
\end{equation}
Second, through the linearity of the individual parameters model in equation (\ref{eq:ind_par}), the posterior distribution $p(\tilde{b}_i|\phi_i;\theta)$ of the $i$th subject is explicit: $p(\tilde{b}_i|\phi_i;\theta)$ is a Gaussian distribution $\mathcal{N}(m(\phi_i,\theta),V(\theta))$ of mean and variance equal to:
\begin{eqnarray}\label{postb}
m(\phi_i,\theta)&=&V(\theta)\left(\Psi^{-1}\sum_{k=1}^K (\phi_{ik}-\beta_k)+\Omega^{-1}\mu\right),\\
V(\theta)&=&(\Omega^{-1}+K\Psi^{-1})^{-1}.\nonumber
\end{eqnarray}

Due to the factorization given in equation (\ref{bayes}),  function $Q$  can be rewritten as:
\begin{eqnarray*}
Q(\theta|\theta')&=&\int\left( \int  L_c(\y,\phi,\tilde{b};\theta)p(\tilde{b}|\phi;\theta')d\tilde{b} \right)p(\phi|\y;\theta')d\phi.
\end{eqnarray*}
Because of the explicit posterior distribution of random effects $\tilde{b}$ given in equation (\ref{postb}), the computation of this conditional expectation can be split into two parts : the  computation of the  integral with respect to the posterior distribution of $\tilde{b}$ which has an analytical form, and the computation of the integral with respect to the posterior distribution of $\phi$ which has no analytical form. Therefore the EM algorithm is split into an explicit direct EM algorithm for the computation of  the first integral and  the use of a stochastic version of the EM algorithm for the computation of the second integral.

Let us detail the explicit computation of the first integral, denoted by  
 $R(\y,\phi,\theta,\theta')$
 $$R(\y,\phi,\theta,\theta')=\int  L_c(\y,\phi,\tilde{b};\theta)p(\tilde{b}|\phi;\theta')d\tilde{b}.$$
This integral has an analytical form.  Indeed, the complete log likelihood $L_c(\y,\phi,\tilde{b};\theta)$ is equal to
\begin{eqnarray*}\label{eq:complete_lik}
L_c(\y,\phi,\tilde{b};\theta)&=&-\frac{1}{2}\sum_{k=1}^{K} \sum_{i=1}^{n}\sum_{j=1}^{n_{ik}}\log(2\pi \sigma^{2}g^2(t_{ijk},\phi_{ik})) -\frac{1}{2}\sum_{k=1}^{K} \sum_{i=1}^{n}\sum_{j=1}^{n_{ik}} \frac{(y_{ijk}-f(t_{ijk},\phi_{ik}))^2}{\sigma^{2}g^2(t_{ijk},\phi_{ik})}\nonumber\\
&&-\frac{nK}{2}\log(2\pi \det\Psi) -\frac{1}{2}\sum_{k=1}^{K}\sum_{i=1}^{n}(\phi_{ik}-\tilde{b}_i-\beta_k)^t \Psi^{-1}(\phi_{ik}-\tilde{b}_i-\beta_k)\\
&&-\frac{n}{2}\log(2\pi \det\Omega) -\frac{1}{2}\sum_{i=1}^{n}(\tilde{b}_i-\mu)^t \Omega^{-1}(\tilde{b}_i-\mu).\nonumber
\end{eqnarray*}
As the posterior distribution $p(\tilde{b}|\phi;\theta)$ is known (equation \ref{postb}), $R(\y,\phi,\theta,\theta')$ is equal to
 \begin{eqnarray}\label{eq:LC}
&&R(\y,\phi,\theta,\theta')=-\frac{1}{2}\sum_{i,j,k} \log(2\pi \sigma^{2}g^2(t_{ijk},\phi_{ik})) \frac{1}{2}\sum_{i,j,k}\frac{(y_{ijk}-f(t_{ijk},\phi_{ik}))^2}{\sigma^{2}g^2(t_{ijk},\phi_{ik})}\\
&& -\frac{nK}{2}\log(2\pi \det\Psi)-\frac{nK}{2}\Psi^{-1/2}V(\theta')\Psi^{-1/2}-\frac{1}{2}\sum_{i,k}(\phi_{ik}-m(\phi_i,\theta')-\beta_k)^t \Psi^{-1}(\phi_{ik}-m(\phi_i,\theta')-\beta_k)\nonumber\\
&&-\frac{n}{2}\log(2\pi \det\Omega) -\frac{n}{2} \Omega^{-1/2}V(\theta')\Omega^{-1/2}-\frac{1}{2}\sum_{i=1}^{n} (m(\phi_i,\theta')-\mu)^t \Omega^{-1}(m(\phi_i,\theta')-\mu).\nonumber
\end{eqnarray}

Therefore $Q$ is reduced to the computation of the second integral under the posterior distribution $p(\phi|\y;\theta)$ as follows:
\begin{equation}\label{eq:Q}
Q(\theta|\theta')=\int R(\y,\phi,\theta,\theta')p(\phi|\y;\theta')d\phi.
\end{equation}
Given the non-linearity of function $f$ with respect to $\phi$, the posterior distribution $p(\phi|\y;\theta')$ has no closed form and the function $Q$ defined by (\ref{eq:Q}) is intractable. Thus we  propose to use the  stochastic version SAEM of the EM algorithm proposed by Delyon et al. 
\cite{del99} which evaluates the integral $Q$ by a stochastic approximation procedure.

Let us detail this SAEM algorithm in the case of two-levels NLMEMs. Let us note that the quantity  $R(\y,\phi,\theta,\theta')$ belongs to the regular curved exponential family, i.e, it can be written as 
\begin{equation}\label{eq:LC_exp}
R(\y,\phi,\theta,\theta')= -\Lambda(\theta)+\langle S(\y,\phi,\theta'),\Phi(\theta)\rangle, 
\end{equation}
where $\left\langle .,.\right\rangle$ is the scalar product, $\Lambda$ and $\Phi$ are two functions twice continuously differentiable on $\Theta$ and $S(\y,\phi,\theta')$ is known as the minimal sufficient statistics of the complete model. Those statistics are detailed later. In this case, the $Q$ function is reduced to 
$$ 
Q(\theta|\theta')=-\Lambda(\theta)+\langle\  \left(\int S(\y,\phi,\theta')p(\phi|\y;\theta')d\phi\right),\ \Phi(\theta)\ \rangle,
$$

In this case, at the $\ell$th iteration, the SAEM algorithm proceeds as follows:
\begin{itemize}
\item Simulation step:  simulation of the missing data $(\phi_i^{(\ell)})_i$ under the
conditional distribution $p (\phi|\y;\widehat{\theta}_\ell)$
\item Stochastic approximation step: computation of a stochastic
approximation $s_{\ell+1}$ of $E\left\lbrack  S(\y,\phi,\widehat{\theta}_{\ell} ) \vert  \y;\widehat{\theta}_{\ell} \right\rbrack=\int S(\y,\phi,\widehat{\theta}_{\ell} )p(\phi|\y;\widehat{\theta}_\ell)d\phi$, using $(\gamma_\ell)_{\ell\geq 0}$, a sequence of positive numbers decreasing to 0:
\begin{equation*}
s_{\ell+1}=s_\ell+\gamma_\ell( S(\y,\phi^{(\ell)},\widehat{\theta}_{\ell} )-s_\ell).
\end{equation*} 
\item Maximization step: update of the estimate $\widehat{\theta}_{\ell+1}$:
$$\widehat{\theta}_{\ell+1}= \arg \max_{\theta \in \Theta} \left(-\Lambda(\theta))+\langle s_{\ell+1},\Phi(\theta))\rangle \right).$$

\end{itemize}

The sufficient statistics of the complete model (\ref{eq:complete_lik})  evaluated during the SA step of the SAEM algorithm are as follows:
\begin{eqnarray*}
s_{1,i,\ell+1}&=&s_{1,i,\ell}+\gamma_\ell\left(\sum_{k=1}^K \phi_{ik}^{(\ell)}-s_{1,i,\ell}\right), \hspace{1em} i=1,\ldots,N, \\
s_{2,k,\ell+1}&=&s_{2,k,\ell}+\gamma_\ell\left( \sum_{i=1}^n\phi_{ik}^{(\ell)}-s_{2,k,\ell}\right), \hspace{1em} k=1,\ldots,K, \\
s_{3,\ell+1}&=&s_{3,\ell}+\gamma_\ell\left( \sum_{i=1}^n m(\phi_i^{(\ell)},\widehat{\theta}_\ell)^t m(\phi_i^{(\ell)},\widehat{\theta}_\ell)-s_{3,\ell}\right),\\
s_{4,\ell+1}&=&s_{4,\ell}+\gamma_\ell\left( \sum_{k=1}^K\sum_{i=1}^n \left(\phi_{ik}^{(\ell)}-m(\phi_i^{(\ell)},\widehat{\theta}_\ell)\right)^t\left(\phi_{ik}^{(\ell)}-m(\phi_i^{(\ell)},\widehat{\theta}_\ell)\right)-s_{4,\ell}\right),\\
s_{5,\ell+1}&=&s_{5,\ell}+\gamma_\ell\left(\sum_{i,j,k}\left(\frac{y_{ijk}-f(t_{ijk},\phi_{ik}^{(\ell)})}{g(t_{ijk},\phi_{ik}^{(\ell)})}\right)^2-s_{5,\ell}\right),
\end{eqnarray*}
The expression of the M step is obtained by derivation of equation (\ref{eq:LC}).
The parameter estimates are as follows:
\begin{eqnarray*}
\widehat{\mu}_{\ell+1}&=& V(\widehat{\theta}_\ell)\widehat{\Psi}_{\ell}^{-1}\left(\frac{1}{n}\sum_{i=1}^ns_{1,i,\ell+1}-\sum_{k=1}^K\widehat{\beta}_{k,\ell}\right)+V(\widehat{\theta}_\ell)\widehat{\Omega}_{\ell}^{-1}\widehat{\mu}_{\ell}  ,\\
\widehat{\beta}_{k,\ell+1}&=& \frac{ s_{2,k,\ell+1}}{n} - \widehat{\mu}_{\ell+1}, \hspace{3em} \text{for } k=2,\ldots,K,\\
\widehat{\Omega}_{\ell+1}&=&V(\widehat{\theta}_\ell)+\frac{s_{3,\ell+1}}{n}-(\widehat{\mu}_{\ell+1})^t\widehat{\mu}_{\ell+1},\\
\widehat{\Psi}_{\ell+1}&=&V(\widehat{\theta}_\ell)+\frac{s_{4,\ell+1}}{nK}-\frac{1}{K}\sum_{k=1}^K (\widehat{\beta}_{k,\ell+1})^t\widehat{\beta}_{k,\ell+1},\\
\widehat{\sigma^2}_{\ell+1}&=&\frac{s_{5,\ell+1}}{\sum_{i=1}^n \sum_{k=1}^K n_{ik}}.
\end{eqnarray*}

Comparing with the classic SAEM algorithm for single-level NLMEMs, the extension of SAEM to the two-levels NLMEMs is finally split into an explicit EM algorithm and a stochastic EM part. Furthermore, it requires the computation of two intermediate quantities (the conditional expectations $m(\phi_i,\theta)$ and variance $V(\theta)$  of the between-subject random effects parameters $b_i$) as well as two additional sufficient statistics ($S_3$ and $S_4$), functions of $m(\phi_i,\theta)$. The M-step differs from the one of the classic SAEM for single-level NLMEMs, especially for the estimation of the variance matrix $\Omega$ and $\Psi$ which uses the  additional quantity  $V(\theta)$.

As proved by \cite{del99,kuh04}, the convergence of the SAEM algorithm is ensured under the following assumption:

\emph{ Assumption} (\textbf{A1}):
\begin{enumerate}
\item  Functions $\Lambda$ and $\Phi$ are twice continuously differentiable on $\Theta$.
\item The log-likelihood  $\log p(\y;\theta)$ is $d$ times differentiable on $\Theta$, where $d$ is the dimension of $S(\y,\phi,\theta')$.
\item Function $\bar{s}$ defined as
$$ \bar{s}(\theta,\theta')=\int S(\y,\phi,\theta')p(\phi|\y;\theta)d\phi $$
is continuously differentiable on $\Theta$ with respect to its first variable.
	\item For all $\ell$ in $\mathbb{N}$, $\gamma_\ell \in [0,1], \sum_{\ell=1}^\infty \gamma_\ell =\infty$ and $\sum_{\ell=1}^\infty \gamma_\ell^2 <\infty$.
\end{enumerate}
For a convenient step sizes sequence $\gamma_\ell$, the assumption (\textbf{A1}) is trivially checked in our model. A choice of step sizes sequence $\gamma_\ell$ is presented in Section \ref{simu}.

However, the simulation step of the SAEM algorithm, which performs the simulation of the non-observed vector $\phi$ under the posterior distribution $p(\phi|\y;\theta)$ cannot be directly performed because the posterior distribution is only known up to a constant. In this case,  Kuhn and Lavielle \cite{kuh05} propose to combine the SAEM algorithm with a Markov Chain Monte Carlo (MCMC) procedure for the simulation step. This version of the SAEM-MCMC algorithm can be used for the estimation of MNLMEMs. The MCMC procedure used in this case is detailed in section \ref{mcmc}.
As proved by \cite{kuh04}, the convergence of the SAEM-MCMC algorithm is ensured under assumption (\textbf{A1}) and the following additional assumption:

\emph{Assumption} (\textbf{A2}):

For any $\theta$ in $\Theta$, we assume that the conditional distribution $p(.|\y;\theta)$ is the unique limiting distribution of a transistion probability $\Pi_{\theta}$, that has the following properties:
\begin{enumerate}


\item For any compact subset $V$ of $\Theta$, there exists a real constant $L$ such that for any $(\theta,\theta')$ in $V^2$
$$
\sup_{\{ \phi, \phi'\} \in \mathcal{E}} \left|\Pi_{\theta}\left(\phi'| \phi \right)-\Pi_{\theta'}\left(\phi'| \phi\right) \right| \leq L\|\theta-\theta'\|.$$
\item The transition probability $\Pi_{\theta}$ supplies an uniformly ergodic chain whose invariant probability is the conditional distribution $p(\phi|\y ; \theta)$, i.e.
$$
\exists K_{\theta} \in \mathbb{R}^+, \mbox{\hspace{1em}}  \exists \rho_{\theta}\in ]0,1[ \mbox{\hspace{1em}}   |  \mbox{\hspace{1em}}  \forall \ell \in \mathbb{N} \mbox{\hspace{1em}} 
\|\Pi^\ell_{\theta}(\cdot|\phi)-p(\cdot,\cdot|\y ; \theta)\|_{TV} \leq C_{\theta} \rho^\ell_{\theta}, 
$$
where $\| \cdot \|_{TV}$ is the total variation norm. Furthermore,
$$ C=\sup_{\theta \in \Theta} C_{\theta} < \infty  \mbox{\hspace{1em} and \hspace{1em}} \rho=\sup_{\theta \in \Theta}  \rho_{\theta} <1.$$

\item Function $S$ is bound on $\mathcal{E}$. 
\end{enumerate}
At iteration $\ell$, the S-step of the SAEM-MCMC algorithm consists thus in simulating $\phi^{(\ell)}$ with the transistion probability $\Pi_{\hat{\theta_{\ell}}}\left(\phi^{(\ell-1)}| d\phi^{(\ell)} \right)$.

 The assumption (\textbf{A2}.2) is the most delicate to check, and it depends on the choice of the MCMC algorithm. This is detailed in the next subsection after presenting the MCMC procedure. 

In practice, the SAEM algorithm being a stochastic algorithm, there exists no deterministic convergence criterion which could be used to stop the iterations of the algorithm as soon as the convergence is reached. Therefore, we recommend to implement the SAEM algorithm with a sufficiently large number of iterations and to graphically check the convergence by plotting the values of the SAEM estimates obtained along iterations versus the iterations. Such a figure is described in Section \ref{simu}.

\subsection{MCMC algorithm for the simulation step}\label{mcmc}

Let us detail the simulation step of the SAEM-MCMC algorithm, which performs the simulation of the missing data $\phi$ through a Markov chain which has $p(\phi|\y;\theta)$ as unique stationary distribution. For subject $i$, by Bayes formula, this conditional distribution is proportional to
$$p(\phi_i|\y_i;\theta)\propto \prod_{k=1}^K p(\y_{ik}|\phi_{ik};\theta)p(\phi_i;\theta).$$

We propose to use a Metropolis-Hastings (M-H) algorithm to simulate this Markov chain. Let us recall the principle of this algorithm. At iteration $r$ of the M-H algorithm, given the current value $\phi_{i}^{(r)}$ of the Markov Chain, the M-H algorithm proceeds as follows:
\begin{enumerate}
	\item Simulate $\phi_i^c$ with a proposal distribution $q(\cdot,\phi_i^{(r)})$
	\item Compute the acceptance probability
	$$\alpha(\phi_i^c,\phi_i^{(r)})=\frac{\prod_{k=1}^K p(\y_{ik}|\phi_{ik}^c;\theta)p(\phi_i^c;\theta)}{\prod_{k=1}^K p(\y_{ik}|\phi_{ik}^{(r)};\theta)p(\phi_i^{(r)};\theta)}\frac{q(\phi_i^c,\phi_i^{(r)})}{q(\phi_i^{(r)},\phi_i^c)} $$
	\item Simulate $u$ with a uniform distribution $\mathcal{U}[0,1]$
	\item Update the Markov chain with
	\begin{equation*}
	\phi_i^{(r +1)}=\left\{
	\begin{array}{ccc}
	\phi_i^c & \text{if} &u\leq\alpha(\phi_i^c,\phi_i^{(r)})\\
		\phi_i^{(r)} & \text{else} &
	\end{array}
\right.
	\end{equation*}
\end{enumerate}
The convergence of the M-H algorithm strongly depends on the choice of the proposal distribution $q$. The convergence is ensured for some proposal distributions such as independent ($q(\cdot,\phi_i^{(r)})$ independent of $\phi_i^{(r)}$) or symmetrical ($q(\cdot,\phi_i^{(r)})=q(\phi_i^{(r)},\cdot)$) proposals \cite{men93}. These proposals are detailed below. Given the dimension of $\phi$, we also consider a Metropolis-Hastings-Within-Gibbs algorithm, combining both Gibbs algorithm and M-H procedure.  The advantage of the Gibbs algorithm is to reduce the multi-dimensional simulation problem to the successive simulations of one-dimension vectors. 
Finally, at iteration $\ell$ of the SAEM algorithm, given the current estimate $\widehat{\theta}_\ell$, we combine the three following proposal transitions:  
\begin{enumerate}
	\item 
	the prior distribution of $\phi_i$, that is the Gaussian distribution $\mathcal{N}(\widehat{\mu}_\ell+\widehat{\beta}_\ell,\widehat{\Gamma}_\ell)$, corresponding to an independent M-H algorithm,
	\item 
	the multidimensional random walks  $\mathcal{N}(\phi_i^{(\ell-1)},\rho \widehat{\Gamma}_\ell)$ (symmetric proposal), where $\rho$ is a scaling value chosen to ensure a satisfactory acceptation rate, namely around 30\% as proposed in \cite{rob05}, 
	\item 
	a succession of $Kp$ unidimensional Gaussian random walks (symmetric proposal), i.e each component of $\phi_i$ is successively updated, leading to a Metropolis-Hastings-Within-Gibbs algorithm, 
\end{enumerate}
where $\widehat{\Gamma}_\ell$ is equal to
\begin{equation*}
\widehat{\Gamma}_\ell=\left(
\begin{array}{ccccc}
\widehat{\Omega}_\ell +\widehat{\Psi}_\ell & \widehat{\Omega}_\ell &\ldots &\widehat{\Omega}_\ell\\
\widehat{\Omega}_\ell & \widehat{\Omega }_\ell +\widehat{\Psi}_\ell&\ddots& \vdots\\
\vdots & \ddots&\ddots&\widehat{\Omega}_\ell\\
\widehat{\Omega}_\ell&\ldots&\widehat{\Omega}_\ell&\widehat{\Omega}_\ell  +\widehat{\Psi}_\ell
\end{array}
\right).
\end{equation*}
Given the proposal distributions, as previously detailed, and using the theoretical convergence results proposed in \cite{men93}, this hybrid Gibbs algorithm converges and generates an uniformly ergodic chain with $p(\phi|\y;\theta)$  as the stationary distribution. 
Consequently, by applying the convergence theorem of Kuhn and Lavielle \cite{kuh05} and under assumptions (A1) and (A2), we prove that the estimate sequence $(\widehat{\theta}_\ell)_{\ell\geq0}$ produced by the extended SAEM algorithm converges towards a (local) maximum of the likelihood $p( \y ;\theta)$.

In practice, the convergence of the MCMC algorithm is difficult to verify. As in Bayesian inference, the only convergence criteria existing for MCMC procedure are  graphical criteria. We have to check if the estimate sequence explores a sufficiency large domain of the Markov chain. A convergence figure is presented and commented in section \ref{simu}.

\subsection{Estimation of the Fisher information matrix and the likelihood}
To perform statistical tests such as Wald test or likelihood ratio test, we propose estimators of the Fisher information matrix and the likelihood, respectively.  
As the Fisher information matrix has no closed form in MNLMEMs, we propose to approximate it by 
the Fisher information matrix of the multi-level linear mixed model deduced from the MNLMEM after linearization of the function $f$ around the conditional expectation of the individual parameters $(E(\phi_{i}|y;\hat{\theta}), 1\leq i \leq n) $. The computation of this linearized Fisher information matrix is direct and does not need any approximation. 

The estimation of the likelihood of the MNLEM is based on an Importance Sampling procedure, as proposed by Samson et al. \cite{sam06} for NLMEMs. The Importance Sampling procedure has been introduced to approximate the integral of the likelihood with a smaller variance than with other Monte Carlo methods. 
In this case, an estimate of the contribution $p(\y_i;\theta)$ of the individual $i$ to the likelihood is
$$\widehat{p(\y_i;\theta)} =  \frac{1}{T} \sum_{t=1}^T \frac{p(\y_i,\phi_i^{(t)};\theta)}{q_i(\phi_i^{(t)})}$$   
where $(\phi_i^{(t)})_{t=1,\ldots,T}$ are simulated using the individual instrumental distribution $q_i$. As an individual instrumental distribution $q_i$, we propose a Gaussian approximation of the individual conditional posterior distribution $p(\phi_i|\y_i;\theta)$.

\section{Simulation study: a PK example}
\label{simu}

\subsection{Simulation settings}

The  objective of this simulation study is to illustrate the main statistical properties of the extended SAEM algorithm (bias, root mean square errors, group comparison tests) and to compare them to the FOCE algorithm. We do not use the AGQ algorithm, since procedure NLMIXED does not succeed, in practice, in estimating such complex variance models.

We use the PK data of orally administered theophyllin to define the population model for the simulation study. These data are classical ones in population pharmacokinetics, often used
for software evaluation \cite{pin00}. We assume that concentrations can be described by a one-compartment model with first order absorption and first order elimination:
\begin{equation*}
f({t,\phi})=\frac{DK_{a}}{VK_{a}-Cl} \left(e^{-\frac{Cl}{V}t}-e^{-K_{a}t} \right) \label{e4}
\end{equation*}
where $D$ is the dose, $V$ is the volume of distribution, $K_{a}$ is the absorption rate constant and $Cl$ is the clearance of the drug elimination. 
These parameters are positive and distributed according to a log-normal distribution. Thus, $\phi$ has the following components:  $\phi=(\log V, \log K_{a}, \log AUC)$, with $AUC=D/Cl$.
We assume identical sampling times for all subjects: for all $i$ in $1,\ldots,n$, $k=1,2$,  $t_{ijk}=t_j$ for $j=1,\ldots, J$.
Additive Gaussian random effects are assumed for each parameter with a diagonal covariance matrix $\Omega$ and a a diagonal covariance matrix $\Psi$.  Let  $\omega^2=(\omega_V^2,\omega_{K_a}^2,\omega_{AUC}^2)$ and $\psi^2=(\psi_V^2,\psi_{K_a}^2,\psi_{AUC}^2)$ denote the  vector of the variances of the random effects. A combined error model is assumed by setting $g(t,\phi)=1+f(t,\phi)$.

\begin{figure}
	\centering
		\includegraphics[width=15cm]{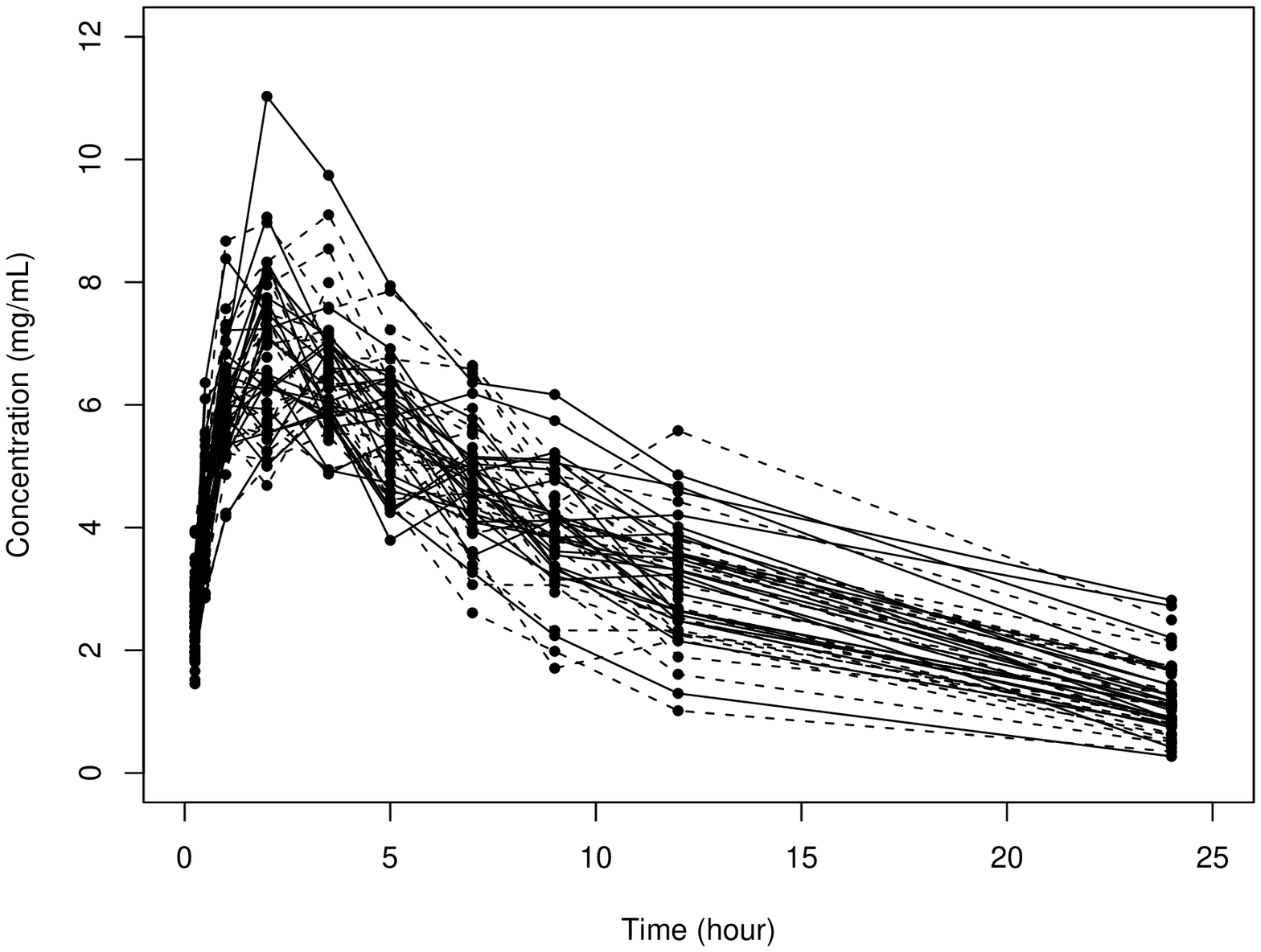}
	\caption{Simulated theophyllin concentration data for 24 subjects during the first period (plain line) and during the second period (dotted line)}
	\label{fig:theoph_data}
\end{figure}

We set the dose for all subjects to the value of 4 mg. For all the parameters, the values are those proposed by Panhard and Mentr\'e \cite{pan05}: $\log V=-0.73$, $\log K_{a}=0.39$ and $\log AUC=4.61$,  $\omega_V^2=0.01$, $\omega_{K_a}^2=0.04$, $\omega_{AUC}^2=0.04$, $\psi_V^2=0.0025$, $\psi_{K_a}^2=0.01$, $\psi_{AUC}^2=0.01$ and $\sigma^2=0.01$. 
We generate  $n=24$ and $n=40$ total numbers of subjects  with $J=10$ blood samples per subject, taken at 15 minutes, 30
minutes, 1, 2, 3.5, 5, 7, 9, 12 and 24 hours after dosing.
The individual data of one simulated trial are displayed in Figure \ref{fig:theoph_data}.

\subsection{Evaluation of estimates}
Our aim is to evaluate and compare the estimates produced  by the extended SAEM algorithm with those  produced by the nlme function of the R software. 
We fit the simulation model and compute the relative bias and relative root mean square error (RMSEs) for each component of $\theta$ from  1000 replications of the two trials described below ($n=24$ and $n=40$ total number of subjects). 

The simulation model includes a treatment effect on all components of $\theta$. We test the null hypothesis $\{\beta_{\log AUC} =0\}$ using the Wald test. We also fit the model where the treatment effect on $\log AUC$ is not estimated, and test the same null hypothesis using the Likelihood Ratio Test (LRT). 

\begin{figure}[h]
   \centering
       \includegraphics[width=1.1\textwidth]{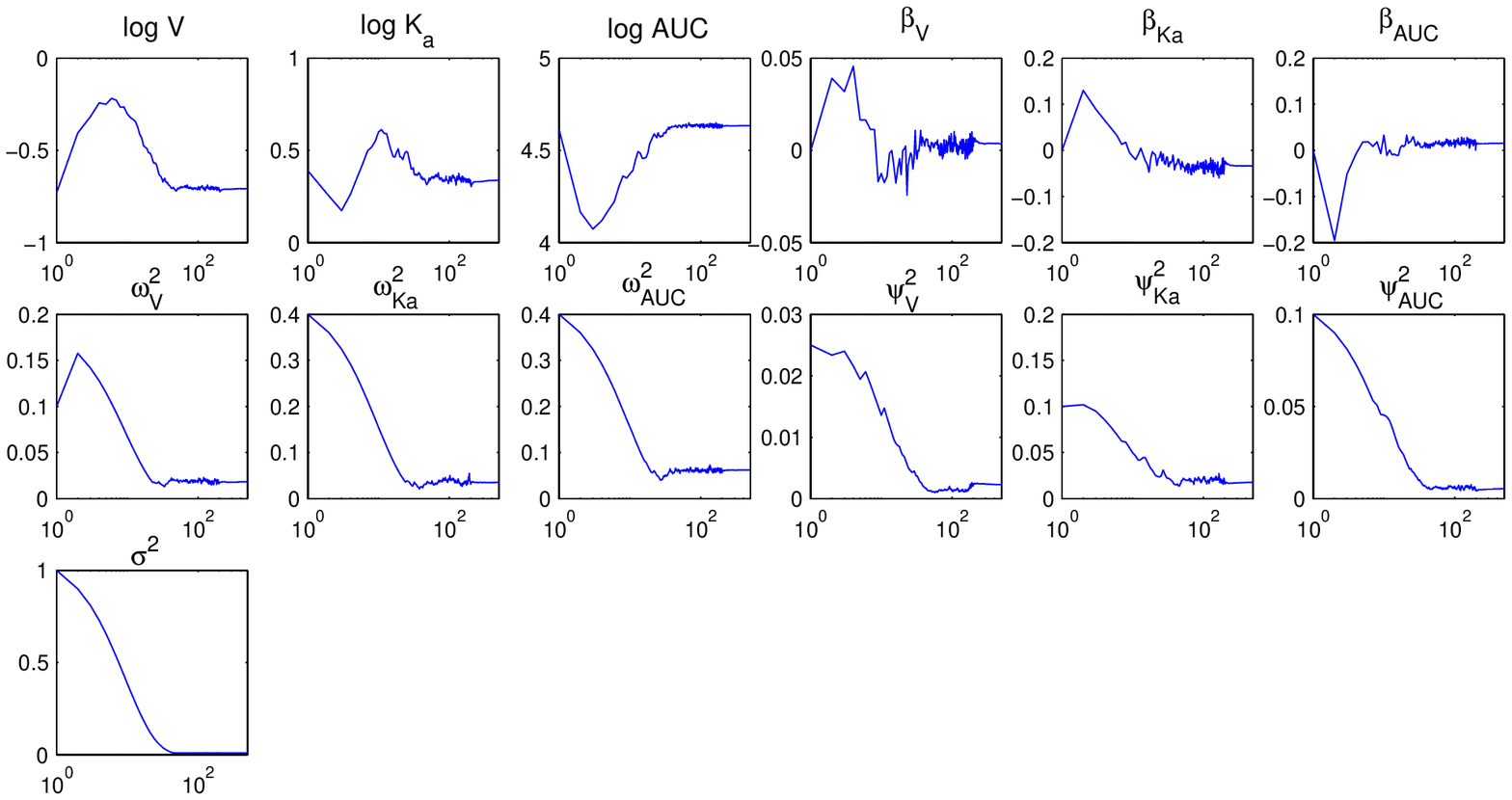}
   \caption{Evolution of the estimates, function of the iteration of SAEM algorithm (with a logarithm scale for the abscis axis).}
   \label{fig:conv_saem}
\end{figure}

The SAEM algorithm is implemented with $500$ iterations. During the first $200$ iterations, a constant step size $\gamma_\ell=1$ is chosen, in order to let the Markov chain explore the parameters domain. Then during the last $300$ iterations, the stochastic approximation scheme is implemented with a step size  equal to $\gamma_\ell=\frac{1}{\ell-200}$ at iteration $\ell$. This choice of step size sequence verifies convergence assumption (\textbf{A1}.1).
The evolution of each SAEM parameter estimates
is plotted  against iterations (logarithmic scale) on Figure
\ref{fig:conv_saem}. During the first iterations of the SAEM algorithm, the estimate sequences explore randomly some neighborhoods of the initial values, through the Markov chain simulation. In particular,  these behaviors are clearly visible for the fixed effect parameters ($\mu$ and $\beta$). After 200 iterations, the estimates converge then rapidly to a neighborhood of the maximum likelihood, due to the stochastic approximation scheme. In this example, the iteration number has been chosen such that the convergence is clearly attained before the last iteration.

The relative bias and RMSEs obtained on the 1000 datasets with $n=24$ and $n=40$ subjects are presented in Table \ref{tab1}.
\begin{table}
\caption{ Relative biases ($\%$) and relative root mean square errors (RMSE) ($\%$) of the estimated parameters evaluated by the extended SAEM algorithm and the FOCE algorithm (nlme function) from 1000 simulated trials.}\label{tab1}
\vspace{0.5em}
\centering
\begin{tabular}{l r r r r c r r r r}
 \hline
 &\multicolumn{4}{c}{n=24 subjects} &&\multicolumn{4}{c}{n=40 subjects}\\
 \hline
 &\multicolumn{2}{c}{Bias} &\multicolumn{2}{c}{RMSE}&&\multicolumn{2}{c}{Bias} &\multicolumn{2}{c}{RMSE}\\
 \hhline{~----~----}
   & SAEM & nlme& SAEM & nlme&& SAEM & nlme& SAEM & nlme\\
 \hline
$V$ &  0.01  &  0.53 &  3.9  &  6.4 & & -0.06&  0.54   &   2.91  & 5.00  \\
$k_a$ &  0.48  &  -1.48 &  14.4  &  24.4&& 0.02&  -3.07   &  10.79   & 18.4   \\
$AUC$ &  -0.08  &  -0.20 &  1.0  & 1.5 &&   -0.11 &  -0.24 &  0.79    & 1.13   \\
$\beta_V$ &  -0.00  &  -0.01 &  3.6  & 3.6& &  -0.05&  -0.06     &2.83 &2.80		   \\
$\beta_{k_a}$ &  -0.73  &  -0.76 &  14.2  &  18.8 & & 0.24& 0.27& 10.73   & 10.60    \\
$\beta_{AUC}$ &  0.02  &  -0.02 &  0.7  & 1.1 & & 0.00& 1.14& 0.57  &  1.14  \\

$\omega^2_{V}$ &  -5.13  &  -5.92 &  38.7  &  38.4 &&   -3.45&  -4.28  & 30.30   & 30.37    \\
$\omega^2_{k_a}$ &  -3.99  &  -7.07 &  42.4  &  41.5 && -3.23  &  -4.63 & 33.49   & 33.32     \\
$\omega^2_{AUC}$ &  -4.88  &  -7.29 &  34.5 & 34.2 & &   -1.51 &  -3.80 & 27.41    & 27.02     \\
$\psi^2_{V}$ &  -8.67  &  -7.29 &  69.4  &  68.5&&     -5.93 &  -4.91 & 58.78  & 57.81     \\
$\psi^2_{k_a}$ &  -10.94  &  -9.09 &  73.5  &  72.0 &&    -7.06 &  -5.17 & 62.00    & 60.60   \\
$\psi^2_{AUC}$ &  -5.37  &  -5.37 &  43.6  & 42.6 &&    -4.92 &  -5.79 & 33.31    & 32.47\\
$\sigma^2$ & -0.33   &-0.10 & 7.7   & 7.7 & &       0.28 &  0.67 &  6.03     & 6.09                  \\
 \hline
\end{tabular}
\end{table}

   The bias and the RMSEs of the fixed effects ($\mu$) are small with the SAEM algorithm and are almost  half of those obtained with nlme (especially the RMSEs). The bias and RMSEs of the unit effect ($\beta$) are small and on the same order with both methods. 
For the between-subject   variability parameters ($\Omega$), the bias are reduced with SAEM, while the RMSEs are of the same order with both methods. For the within-subject variability parameters ($\Psi$), the bias and the RMSEs are satisfactory, and on the same order with both methods. The bias and RMSE for $\sigma^2$ are small and satisfactory for both methods.

The type I error of the Wald test and of the LRT are evaluated on the same 1000 datasets. For $n=24$, the type I errors are 6.0\% and 6.5\% for SAEM and nlme, respectively, for the Wald test, and 4.6\% and 5.6\% for SAEM and nlme, respectively, for the LRT. For $n=40$, the type I error are 5.6\% and 5.4\% for SAEM and nlme, respectively, for the Wald test, and 5.8\% and 5.2\% for SAEM and nlme, respectively, for the LRT.

\section{Application to the population pharmacokinetics of atazanavir with tenofovir}\label{ataza}

\subsection{Study population: ANRS 107 - Puzzle 2 study}

The Puzzle 2 - ANRS  107 trial was a randomized open-label, multiple-dose study supported by the French Agence Nationale de Recherche sur le Sida (ANRS) with HIV-infected patients in treatment failure with their antiretroviral therapy.  
Patients were randomized to receive for the first two weeks either their unchanged treatment with PIs and nucleoside reverse transcriptase inhibitors (NRTIs) (group 1) or unchanged treatment with NRTIs in combination with atazanavir (300 mg QD)  plus ritonavir (100 mg QD) as a substitute for the failing PI therapy (group 2). From week 3 (day 15) to week 26, patients from either group switched to atazanavir (300 mg QD) plus ritonavir (100 mg QD) plus tenofovir DF at 300 mg QD and NRTIs selected according to the baseline reverse transcriptase genotype of the HIV isolated in each patient. 
 
In this paper, we analyze concentration data obtained from 10 patients  from group 2 who were included and measured at 1, 2, 3, 5, 8, and 24 h after administering drug during each treatment period. Those exact dosing intervals were recorded. 
  The objective of the substudy was to measure the pharmacokinetic 
 parameters of atazanavir (administered with ritonavir) either before (day 14 [week 2]) or after (day 42 [week 6]) initiation of tenofovir DF in HIV-infected patients in order to detect pharmacokinetic interactions of tenofovir on atazanavir. Data of this substudy were analyzed using a nonlinear mixed effect model by Panhard et al \cite{pan07} and a significant effect of the co-administration of tenofovir on the pharmacokinetic parameters of atazanavir was found using the nlme function of the Splus software. 

The aim of the present analysis is to evaluate the effect of tenofovir on the PK parameters of atazanavir using the SAEM algorithm and the Wald test described in section 4.

We use the one-compartment model with zero-order absorption proposed by Panhard et al. \cite{pan07} to describe atazanavir concentrations:
\begin{displaymath}
f(t,\phi)=\frac{FD}{T_{a}Cl}\left( (1-e^{-\frac{Cl}{V}t})\mathbbm{1}_{t<T_{a}}
	 + \frac{e^{-\frac{Cl}{V}\tau\mathbbm{1}_{t<T_{a}}}(1-e^{-\frac{Cl}{V}T_{a}})e^{-\frac{Cl}{V}(t-T_{a})}}{(1-e^{-\frac{Cl}{V}\tau})}    \right)
\end{displaymath}
with $F$ the bioavailability, $V$ the volume of distribution of atazanavir, ($T_{a}$) the  absorption duration,  $Cl$ the elimination clearance and $\tau$ the dosing interval (24 hours until the PK visit). The vector of the logarithm of the identifiable parameters is $\phi=(\log(V/F), \log(T_a)$, $\log(AUC))$.
Data of both treatment periods are simultaneously analyzed using a NLMEM with two levels of variability (the between-patient and  within-patient variabilities) on each PK parameter. A treatment effect is also estimated for each PK parameter, and a homoscedastic error model is used.

\subsection{Results}
\begin{figure}[h]
		\includegraphics[width=18cm]{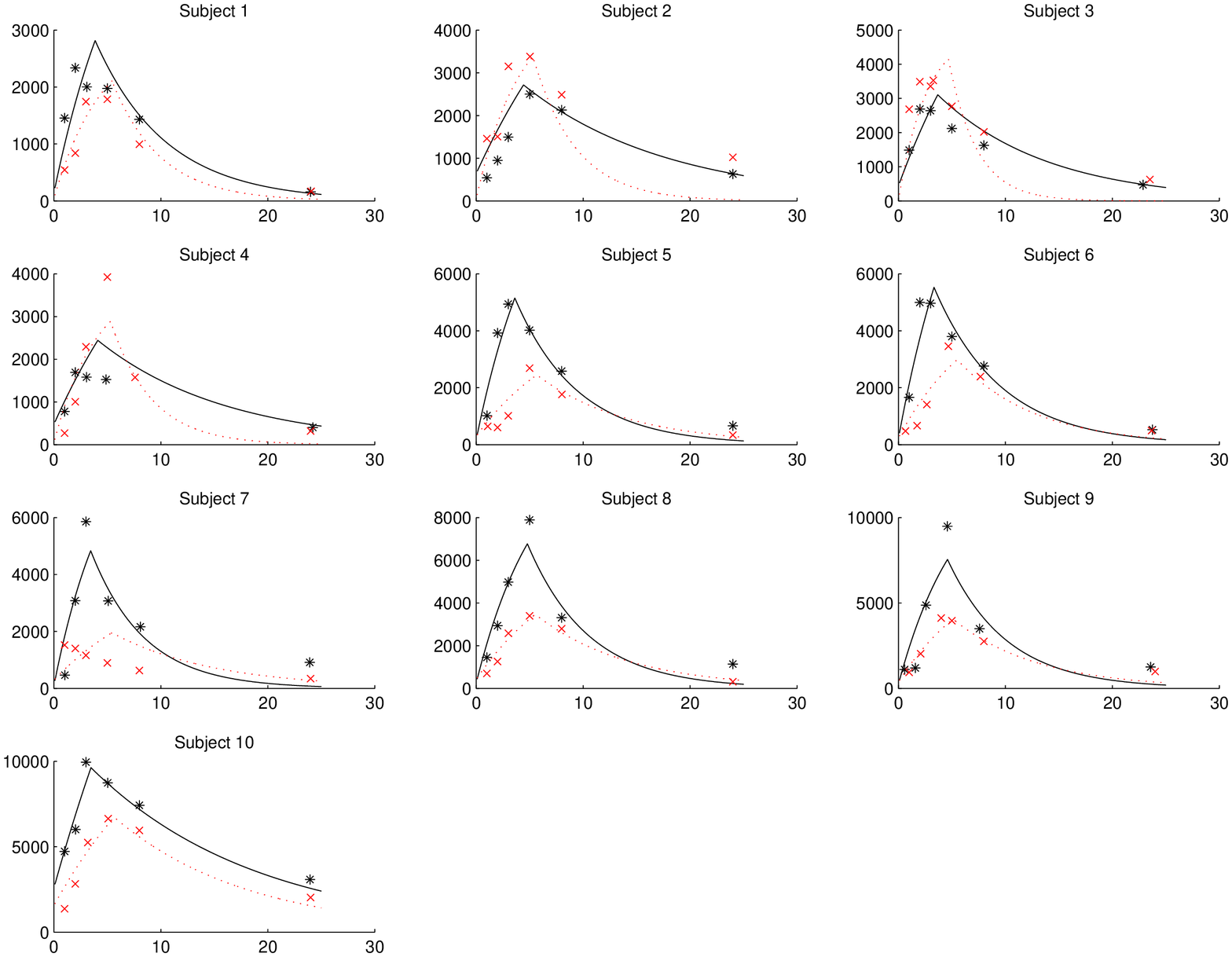}
   \caption{\label{fig1}Individual concentrations and individual predicted curves  for the pharmacokinetics of atazanavir in 10 subjects: x and $*$, observations with and without tenofovir, respectively; dotted and plain line, individual predictions of the atazanavir pharmacokinetics with and without tenofovir, respectively.} 
\end{figure}

Concentrations versus time are displayed in Figure \ref{fig1}.
The SAEM algorithm succeeds in the estimation of all the parameters. The resulting parameters estimates are displayed in table \ref{table_ataza}. The SAEM algorithm estimates the $AUC$ between-patient variability   and the $V/F$ and $T_a$ within-patient variabilities to 0.48, 0.69 and 0.19, respectively. The three other variances are estimated to 0.  
 
\begin{table}
\begin{center}
	\caption{Pharmacokinetic parameters of atazanavir (estimate and SE (\%))  estimated with the SAEM }\label{table_ataza} 
	\vspace{1em}
		\begin{tabular}{l r r }
	 \hline

Parameters & Estimate & SE (\%) \\
 \hline
 $\log(V/F)$ (L)&    4.01 & 5.79  \\
 $\log(T_a)$ (h)&   1.36  &  6.72   \\
 $\log(AUC)$ (ng.mL$^{-1}$.h)&   10.67  &  1.61  \\
 $\beta_{V/F}$&0.12 &   267.43  \\
$\beta_{T_a}$ & 0.33 &  45.03   \\
$\beta_{AUC}$ &  -0.38 &  25.31    \\
 
$\omega_{V/F}$ &0 &  - \\
$\omega_{T_a}$ &  0  & -   \\
$\omega_{AUC}$ &  0.48&  25.48  \\

$\psi_{V/F}$ &0.55 &28.30 \\
$\psi_{T_a}$ &  0.16 &  35.76   \\
$\psi_{AUC}$ &  0 &  -    \\
   $\sigma$ (ng.mL$^{-1}$) &  732.29   & 8.40 \\
 \hline
		\end{tabular}
		\end{center}

\end{table}
A significant effect of co-medication with tenofovir is found on $\log(AUC)$ (p=0.00015)  with the Wald test based on the SAEM algorithm.

\begin{figure}

		\includegraphics[width=18cm]{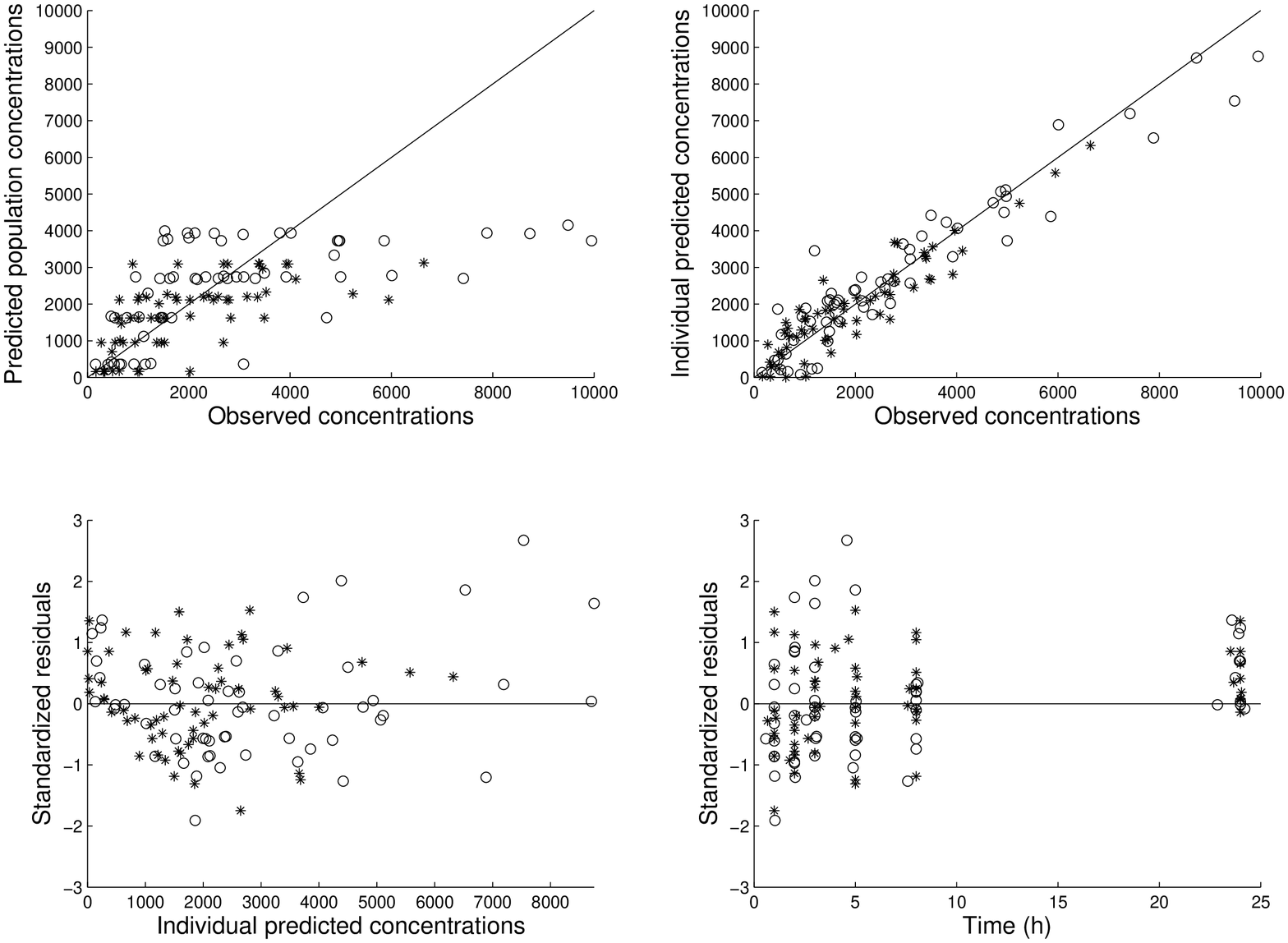}
	
	\caption{Goodness-of-fit plots for atazanavir final population PK model: population (a) and individual
(b) predicted concentrations (in ng/mL) versus observed concentrations (in ng/mL), standardized residuals
versus predicted concentrations (in ng/mL) (c) and versus time (in hours) (d).}\label{gof_article}
\end{figure}

The individual prediction curves for the two periods are overlaid on the concentration data on Figure \ref{fig1} for  10 patients. The goodness-of-fit plots (population and individual predicted concentrations versus
observed concentrations; standardized residuals versus predicted concentrations and versus time)
are judged satisfactory, and are displayed in Figure \ref{gof_article}.

\section{Discussion}\label{disc}

The main original element of this study is the development of the SAEM algorithm for two-levels non-linear mixed
effects models. We extend the SAEM algorithm developed by Kuhn and Lavielle \cite{kuh05}, which was not yet adapted to the case of MNLMEMs with two levels of random effects. This algorithm will be implemented in the 3.1 version of the monolix software, freely available on the following website: http://monolix.org. The two levels of random effects are the between-subject variance and the within-subject (or between-unit) variance, with $N$ subjects and $K$ units, with no restriction on $N$ or $K$. We show that the SAEM algorithm is split into two parts: an explicit EM algorithm and a stochastic EM part. The integration of the term $p(b|\phi;\theta)$ in the likelihood results  in the derivation of two additional sufficient statistics compared to the original algorithm. Furthermore it uses two intermediate quantities, the conditional expectations and variance  of the between-subject random effects parameters $b$. The addition of higher levels of variability would therefore require other extensions of the algorithm.
 
The convergence of the algorithm is monitored from a graphical criterion, as shown in Fig1. An automatic implementation of that stopping criterion to optimize both the number of iterations and the stochastic approximation step should be considered in  future work and extension of the Monolix software. 

The simulation study illustrates the accuracy of our approach. We show that the bias and RMSEs obtained by the extended SAEM algorithm are satisfactory for all parameters. The bias are reduced compared to those obtained with the FOCE algorithm implemented in the nlme function of the R software. The bias are especially divided by two for the fixed effects parameters with SAEM. Furthermore, whereas the nlme implementation of the FOCE algorithm does not always converge with both between- and within-patient variability on all parameters, the extended SAEM algorithm does. We develop the tests for a difference between the units, and we obtain type I errors close to the expected 5\% for the Wald test and the LRT.

The analysis of the pharmacokinetics of atazanavir with tenofovir in the Puzzle 2 - ANRS 107 trial also demonstrates the ability of the extended SAEM algorithm to detect treatment interaction on a real data set. When testing for an interaction of tenofovir on the PK of atazanavir, the impact of tenofovir on the absorption of atazanavir is confirmed; more precisely, a decrease of the AUC of atazanavir as shown by Panhard et al. \cite{pan07} is found.

We compare the extension of SAEM to the FOCE algorithm, that is the most popular method in population pharmacokinetics, which is one of the largest application fields of NLMEM. We try to use the NLMIXED procedure of SAS implementing Gaussian quadrature. However, 
procedure NLMIXED does not succeed, in practice, in estimating such complex variance models, on our simulated data, neither on the atazanavir dataset.
The next step is a comparison with a Bayesian estimation of the parameters using Winbugs, which is beyond the scope of this paper. 

The next ambitious development would be an extension of the algorithm to the case of MNLMEMs with more than two levels of random effects, in order to analyze, for instance, genetic data where more than one generation of parents are taken into account. However, it would be difficult to develop a general estimation method since it strongly depends on the relation (linear or not) of the different levels of random effects.

To conclude, the extended SAEM algorithm combines the statistical properties of an exact method together with a high computational efficiency. We thus recommend the use of this method in MNLMEMs.

\section*{Acknowledgments}
The authors are grateful to Marc Lavielle
for his supportive advice and help. 
The authors  would like to thank the principal investigator (Dr Christophe Piketti), the pharmacology coordinator (Dr Anne-Marie Taburet) and  the scientific committee of the Puzzle 2 - ANRS  107 trial for giving us access to the concentration measurements of the patients.\\

During this work, Xavi\`ere Panhard was supported by a grant from the Agence Nationale de la Recherche (France).

  \bibliography{complete_bib}

\newpage
\appendix
\section{Index of notations}

\noindent\textbf{Model notations}\\
$i$ $(i=1,\cdots,n)$: index of subject \\
$j$ $(j=1,\cdots,n_{ik})$: index of measurement of subject $i$ for unit $k$\\
$k$ $(k=1,\ldots,K)$: index of unit\\
$t_{ijk}$: measurement time in unit $k$ for subject $i$ and measurement $j$\\
$y_{ijk}$: observation in unit $k$ for subject $i$ at time$t_{ijk}$\\
$\y=(y_{ijk})_{ijk}$: vector of the observations in the $K$ units for all the $n$ subject \\
$f$ and $g$: non-linear functions linking observations to sampling times\\
$\phi_{ik}$: $p$-vector of the parameters of subject $i$ for unit $k$\\
$\phi_i=(\phi_{i1},\ldots,\phi_{iK})$: $pK$-vector of individual parameter of subject $i$\\
$\phi=(\phi_{ik})_{i=1,\ldots,n, k=1,\ldots,K}$\\
$\mu$: $p$-vector of the mean of the individual parameters for $k=1$\\
$\beta_{k}$: effect of the $k$th unit in comparison to this first unit\\
$\beta=(\beta_1,\ldots,\beta_K)$: vector of the unit effects\\
$b_{i}$: random effect of size $p$ of subject $i$\\
$\tilde{b}_i:=\mu+b_{i}$
$\tilde{b}=(\tilde{b}_1,\ldots,\tilde{b}_n)$
$c_{ik}$: random effect of size $p$  of subject $i$ and unit $k$\\
$\varepsilon_{ijk}$: measurement error\\
$\Omega$: $p\times p$ between-subject covariance matrix\\
$\Psi$: $p\times p$ within-subject covariance matrix\\
$\Gamma$: $pK\times pK$ covariance matrix of the individual parameters $\phi_i$ $(i=1,\cdots,n)$\\
$\sigma^2$: variance of the measurement error\\
$\theta=(\mu,\beta,\Omega,\Psi,\sigma^2)$: vector of all the parameters\\

\noindent\textbf{Algorithm notations}\\
$p(\y,\phi,\tilde{b};\theta)$: likelihood of the complete data\\
$L_c(\y,\phi,\tilde{b};\theta)$: log-likelihood of the complete data\\
$p(\phi,\tilde{b}|y;\theta')$: posterior distribution of $(\phi,\tilde{b})$ given $(y;\theta')$\\
$p(\phi|y;\theta')$: posterior distribution of $\phi$ given $(y;\theta')$\\
$p(\tilde{b}|\phi;\theta)$: posterior distribution of $\tilde{b}$ given $(\phi;\theta)$\\
$m(\phi,\theta)$: mean of $p(\tilde{b}|\phi;\theta)$\\
$V(\theta)$: variance of $p(\tilde{b}|\phi;\theta)$\\
$Q(\theta|\theta'):=E(L_c(\y,\phi,\tilde{b};\theta)|\y;\theta'))$\\
$R(\y,\phi,\theta,\theta'):=\int  L_c(\y,\phi,\tilde{b};\theta)p(\tilde{b}|\phi;\theta')d\tilde{b}$\\
$\Lambda(\theta,\theta')$ and $\Phi(\theta,\theta')$: functions of $\Theta\times \Theta$\\
$\ell$: iteration number of the SAEM algorithm\\
$\phi_i^{(\ell)}$:  missing data simulated at iteration $\ell$ \\
$S(\y,\phi)$: minimal sufficient statistics of the complete model of dimension $d$\\
$s_{\ell+1}$: stochastic approximation of $E\left\lbrack  S(\y,\phi) \vert  \widehat{\theta}_{\ell} \right\rbrack$\\
$(\gamma_\ell)_{\ell\geq 0}$: step sizes sequence\\
$\Pi_{\theta}$: transition probability\\
$q(\phi_i^c,\phi_i^{(\ell)})$: proposal distribution of the Metropolis-Hastings (M-H) algorithm\\
$\phi_i^c$: candidate simulated using $q(\phi_i^c,\phi_i^{(\ell)})$\\
$\alpha(\phi_i^c,\phi_i^{(\ell)})$: acceptance probability of the M-H algorithm\\
$u$: scalar sample generated with a uniform distribution $\mathcal{U}[0,1]$\\
$C_{\theta}$ and $\rho_{\theta}$: constants involved in the proof of uniform ergodicity of the Markov Chain\\
$q_i$: instrumental distribution used for the estimation of $p(\y_i;\theta)$ by Importance Sampling\\
$T$: number of simulated set of parameters in Importance Sampling\\

\noindent\textbf{PK example notations}\\
$D$: drug dose\\
$V$: volume of distribution of the drug (in liters)\\
$K_a$: absorption rate constant (in hours$^{-1}$)\\
$T_a$: absorption duration (in hours)\\
$Cl$: clearance of elimination (in liters.hours$^{-1}$)\\

\end{document}